\documentclass[journal]{IEEEtran}

\usepackage{fancyvrb}
\usepackage{xurl}
\usepackage{booktabs}
\usepackage{url}
\usepackage{amsmath,amsfonts}
\usepackage{algorithmic}
\usepackage{graphicx}
\usepackage{textcomp}

\usepackage{xspace}
\usepackage{subcaption}
\usepackage{tabularx}
\usepackage{amsmath}
\usepackage{balance}
\usepackage{xcolor}
\usepackage{framed}
\usepackage{chngpage}
\usepackage[many]{tcolorbox}

\usepackage[table,xcdraw]{xcolor}

\usepackage{graphicx}

\usepackage{pifont}

\usepackage{enumitem}

\usepackage{xfp} 

\usepackage{tikz} 

\newcommand{\rqa}{$RQ_1$}
\newcommand{\rqb}{$RQ_2$}
\newcommand{\rqc}{$RQ_3$}

\newcommand{\rqaa}{What Is the Relationship Between the Number of Cumulative Fuzzing Sessions and the Bug Detection Rate?}
\newcommand{\rqbb}{Is Coverage Correlated with the Cumulative Number of Fuzzing Sessions?}
\newcommand{\rqcc}{To What Extent Does Increasing Coverage Through Fuzzing Make it Easier to Detect Fuzzing Bugs?}

\newcommand{\rqA}{\rqa: \rqaa}
\newcommand{\rqB}{\rqb: \rqbb}
\newcommand{\rqC}{\rqc: \rqcc}

\newcommand{\ossFuzzProjectDate}{January 7th 2025\xspace}
\newcommand{\ossFuzzProjectNumber}{1,288\xspace}            

\newcommand{\targetProjectDate}{January 7th 2025\xspace}
\newcommand{\targetProjectNumber}{878\xspace}               
\newcommand{\targetProjectRate}{68\%\xspace}                

\newcommand{\TotalFuzzingSession}{approximately 1.12 million\xspace}        

\newcommand{\gotReportProject}{1,201\xspace}                                
\newcommand{\gotReportNumber}{72,660\xspace}                                

\newcommand{\FixedReportProject}{1,125\xspace}                              
\newcommand{\FixedReportNumber}{56,173\xspace}                              %

\newcommand{\ProjectReportProject}{808\xspace}                              
\newcommand{\ProjectReportNumber}{49,470\xspace}                            %

\newcommand{\totalBuildNumber}{3.95 million\xspace}
\newcommand{\buildStartDate}{March 2017\xspace}
\newcommand{\buildEndDate}{January 2025\xspace}

\newcommand{\totalCoverageNumber}{1.04 million\xspace}                      

\newcommand{\connectVulnsAndBuilds}{43,170\xspace}                      
\newcommand{\connectVulnsAndBuildsPercent}{87.3}                        
\newcommand{\cantConnectVulnsAndBuildsPercent}{12.7}                    

\newcommand{\connectVulnsAndCoverageNumber}{5,465\xspace}               

\newcommand{\RQoneMaxFuzzing}{2,263rd\xspace}                           

\newcommand{\RQtwoMaxCoverage}{2,092nd\xspace}                          

\newcommand{\RQoneFirstDetectionRate}{36\%\xspace}                      
\newcommand{\RQoneSecondDetectionRate}{20\%\xspace}                     
\newcommand{\RQoneDecreaseCount}{26th\xspace}                           
\newcommand{\RQoneDecreaseDetectionRate}{4.90}                             

\newcommand{\RQoneQmedian}{2.19\%\xspace}

\newcommand{\RQtwoInitialPhaseStart}{1\xspace}
\newcommand{\RQtwoInitialPhaseEnd}{200\xspace}
\newcommand{\RQtwoInitialPhase}{\RQtwoInitialPhaseStart-\RQtwoInitialPhaseEnd}

\newcommand{\RQtwoCorr}{0.962\xspace}   
\newcommand{\RQtwoMedianCorrEachProject}{0.006\xspace}   

\newcommand{\RQtwoExampleProjectName}{mbedtls\xspace}
\newcommand{\RQtwoExampleProjectCount}{2,247\xspace}
\newcommand{\RQtwoExampleProjectCorr}{-0.99\xspace}

\newcommand{\RQthreeTotalLine}{14,913\xspace}   %

\newcommand{\RQthreeDetectedIRQsFirst}{-0.01\%\xspace}   %
\newcommand{\RQthreeDetectedIRQsEnd}{0.15\%\xspace}   %

\newcommand{\defect}{fuzzing bug\xspace}
\newcommand{\defects}{fuzzing bugs\xspace}

\definecolor{darkgreen}{rgb}{0, 0.5, 0} 
\definecolor{whitesmoke}{rgb}{0.99, 0.99, 0.99} 

\def\Underline{\setbox0\hbox\bgroup\let\\\endUnderline}
\def\endUnderline{\vphantom{y}\egroup\smash{\underline{\box0}}\\}
\def\|{\verb|}

\newcommand{\ie}{\textit{i.e.,}\xspace}
\newcommand{\eg}{\textit{e.g.,}\xspace}

\newcommand{\etal}{\xspace\textit{et al.}\xspace}

\newcounter{findings_no}

\usepackage{listings}

\definecolor{backcolour}{rgb}{0.95,0.95,0.92}
\lstdefinelanguage{diff}{
  morecomment=**[f][\color{red}]{-},         
  morecomment=**[f][\color{darkgreen}]{+},       
  moredelim=**[is][\bfseries]{@@}{@@},
}
\definecolor{backcolour}{rgb}{0.95,0.95,0.92}
\lstdefinelanguage{commit}{ 
  breakindent = 0pt,
  numbers=none,
  backgroundcolor=\color{white},
  frame=single,
  xleftmargin=3.5em,
  numbersep=0em,
  xrightmargin=1.5em,
}

\usepackage[many]{tcolorbox}  
\tcbuselibrary{listings,breakable}
\definecolor{main}{HTML}{D0D3D4}    
\definecolor{sub}{HTML}{D0D3D4}     
\newtcolorbox{dbox}{
    left=2pt,right=2pt,top=2pt,bottom=2pt,
    enhanced, 
    boxrule = 0pt,
    enlarge top by=5pt,
    enlarge bottom by=3pt,
  }
\tcbset{
    sharp corners,
    before skip = 0.1cm,    
    after skip = 0.01cm      
}

\DeclareUnicodeCharacter{2212}{-}

\ifCLASSINFOpdf

\else

\fi

\begin{document}

\title{Large-Scale Empirical Analysis of Continuous Fuzzing: Insights from 1 Million Fuzzing Sessions}

\author{Tatsuya Shirai,
        Olivier Nourry,
        Yutaro Kashiwa,
        Kenji Fujiwara,
        Yasutaka Kamei,
        and~Hajimu Iida
\thanks{T. Shirai, Y. Kashiwa, and H. Iida are with Nara Institute of Science and Technology, Nara, 630-0192, Japan (e-mail: \{shirai.tatsuya.sp1, yutaro.kashiwa, iida\}@naist.ac.jp).}
\thanks{O. Nourry is with Osaka University, Suita, 565-0871, Japan (e-mail:nourry@ist.osaka-u.ac.jp).}
\thanks{Y. Kamei is with Kyushu University, Fukuoka, 819-0395, Japan (e-mail: kamei@ait.kyushu-u.ac.jp).}
\thanks{K. Fujiwara is with Nara Women's University, Nara, 630-8263, Japan (e-mail:kenjif@ics.nara-wu.ac.jp).}
\thanks{Manuscript received April 1, 2025; revised August 26, 2025.}}

\markboth{IEEE TRANSACTIONS ON SOFTWARE ENGINEERING,~Vol.~14, No.~8, August~2025}%
{Shell \MakeLowercase{\textit{et al.}}: Bare Demo of IEEEtran.cls for IEEE Journals}

\newcommand\submittedtext{%
  \footnotesize This work has been submitted to the IEEE for possible publication. Copyright may be transferred without notice, after which this version may no longer be accessible.}

\newcommand\submittednotice{%
\begin{tikzpicture}[remember picture,overlay]
\node[anchor=south,yshift=10pt] at (current page.south) {\fbox{\parbox{\dimexpr0.65\textwidth-\fboxsep-\fboxrule\relax}{\submittedtext}}};
\end{tikzpicture}%
}
\maketitle

\submittednotice

\begin{abstract}
Software vulnerabilities are constantly being reported and exploited in software products, causing significant impacts on society. In recent years, the main approach to vulnerability detection, fuzzing, has been integrated into the continuous integration process to run in short and frequent cycles. This continuous fuzzing allows for fast identification and remediation of vulnerabilities during the development process. Despite adoption by thousands of projects, however, it is unclear how continuous fuzzing contributes to vulnerability detection.

This study aims to elucidate the role of continuous fuzzing in vulnerability detection. Specifically, we investigate the coverage and the total number of fuzzing sessions when \defects are discovered. We collect issue reports, coverage reports, and fuzzing logs from OSS-Fuzz, an online service provided by Google that performs fuzzing during continuous integration. Through an empirical study of a total of \TotalFuzzingSession fuzzing sessions from \targetProjectNumber projects participating in OSS-Fuzz, we reveal that (i) a substantial number of \defects exist prior to the integration of continuous fuzzing, leading to a high detection rate in the early stages; (ii) code coverage continues to increase as continuous fuzzing progresses; and (iii) changes in coverage contribute to the detection of \defects. 
This study provides empirical insights into how continuous fuzzing contributes to \defect detection, offering practical implications for future strategies and tool development in continuous fuzzing.
\end{abstract}


\IEEEpeerreviewmaketitle

\section{Introduction}\label{sec:introduction}

\IEEEPARstart{I}{n} recent years, a significant number of vulnerabilities have impacted the open-source software ecosystem~\cite{mend2021state}. With over 96\% of commercial code making use of open-source software (OSS)~\cite{DBLP:journals/tse/Murphy-HillJSSP21}, any vulnerability present in OSS has the potential to spread across the world. The risk and impacts of vulnerability propagation was clearly demonstrated by vulnerabilities such as Heartbleed~\cite{heartbleed_nvd}, Log4Shell~\cite{log4shell_nvd}, and RegreSSHion~\cite{regresshion_nvd}. The Heartbleed vulnerability exposed security flaws in approximately 
2.0 million HTTPS servers~\cite{DBLP:conf/imc/DurumericKAHBLWABPP14} and RegreSSHion allowed arbitrary code execution with root privileges on an estimated 14 million servers~\cite{regresshion_impact}.

\begin{figure*}[t]
  \centering
  \includegraphics[width=0.75\textwidth]{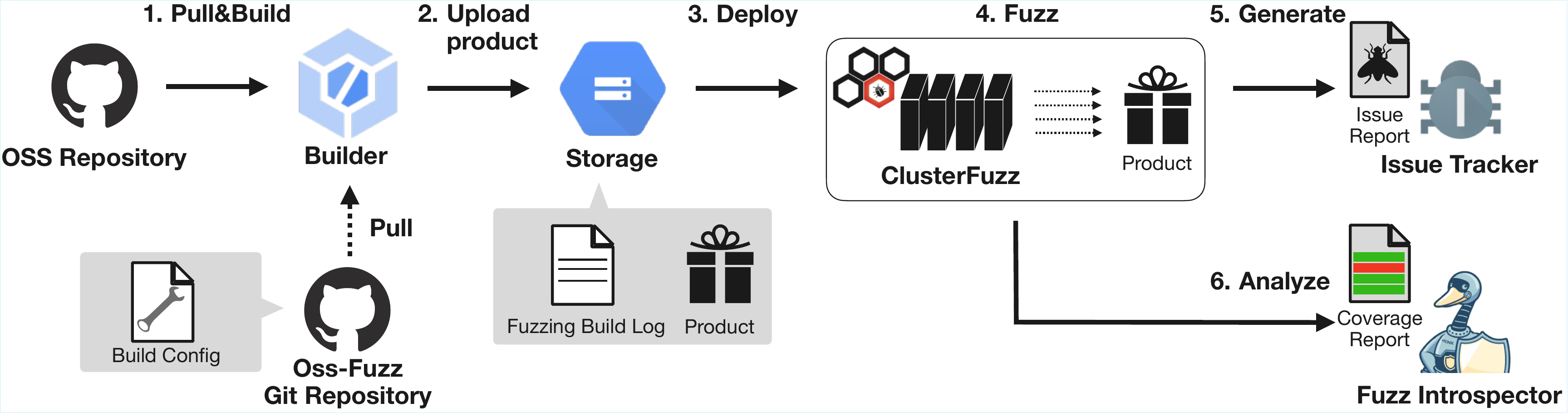}
  \caption{Architecture of OSS-Fuzz}
  \label{fig:oss_fuzz_process}
\end{figure*}

The current state-of-the-art automated approach to detect software vulnerabilities is fuzz testing (fuzzing)~\cite{DBLP:journals/compsec/ChenCMWGL18}. It works by generating large amounts of testing inputs to trigger unintended behavior such as simple crashes, to memory overflows which can be exploited to gain elevated privileges. There are three types of fuzzing strategies: black-box fuzzing, white-box fuzzing and grey-box fuzzing. Black-box fuzzing works by randomly sending large amounts of inputs to the target software. For white-box and grey-box fuzzing, the fuzzer is given knowledge (full knowledge for white-box, partial knowledge for grey-box) about the codebase so that it can send its inputs to specific lines of code.

As of 2025, most modern state-of-the-art fuzzers such as AFL++ and libfuzzer are greybox fuzzers and follow a coverage-guided fuzzing approach (CGF)~\cite{american76:online}. CGF refers to a fuzzing strategy where the fuzzer tries to maximize the coverage (\ie the number of source code lines executed) during fuzzing to increase the chances of finding vulnerabilities. 

Several studies investigated the effectiveness of coverage-based fuzzing. In their study on the reliability of coverage-guided fuzzing, B{\"{o}}hme\etal\cite{DBLP:conf/icse/BohmeSM22} found that there is a strong correlation between code coverage and the number of detected fuzzing bugs (\ie bugs identified by fuzzing). 
Liyanage~\textit{et al.}~\cite{DBLP:conf/icse/LiyanageBTL23} conducted another study where they fuzzed continuously for seven days to study the evolution of code coverage over time. Their results indicate that fuzzing coverage increases logarithmically over time, meaning that long fuzzing sessions give diminishing return over time. 

Despite many studies reporting the efficacy of fuzzing, some studies investigated the current limitations of fuzzing and its pitfalls. For example, Nourry\etal~\cite{DBLP:journals/tosem/NourryKLBLK24} conducted a survey with fuzzing experts to investigate the challenges of fuzzing. Their results revealed several challenges related to the configuration, compilation, and usage of fuzzers. Based on their results, they concluded that the high barrier of entry of fuzzing hinders the adoption of fuzz testing by non-experts (\ie average developers).

To address the complexity and difficulty of adopting fuzzing, external fuzzing providers have gained significant popularity in recent years. As of 2025, the main provider, OSS-Fuzz, is developed and maintained by Google to fuzz critical open-source software systems. One of the major reasons behind the popularity of OSS-Fuzz is its implementation of Continuous Fuzzing through seamless integration within the continuous integration (CI) process~\cite{CI_Fuzz_CLI}. By executing fuzzing periodically and automatically, OSS-Fuzz can provide continuous fuzzing to participating software systems, making early detection of fuzzing bugs more likely.

Traditional fuzzing typically involves executing a fuzzer on a fixed version of the software over an extended period. An industry report suggests that effective fuzzing often spans at least one week, with 30 days considered ideal to search for fuzzing bugs~\cite{Long_Fuzzing_Ideal_Time}. In contrast, continuous fuzzing adopts a different strategy by conducting brief but frequent fuzzing sessions, aiming to identify fuzzing bugs in near real-time as the software evolves. In their study on continuous fuzzing, Klooster\etal~\cite{DBLP:conf/icse/KloosterTBHB23} showed that short sessions of 15 minutes can also be effective to detect critical bugs. In 2021, Zhu\etal~\cite{DBLP:conf/ccs/ZhuB21} conducted a study on grey-box fuzzing and found that 77\% of the fuzzing-related bugs were introduced by recent code changes. Continuous fuzzing provides the ability to fuzz quickly and frequently, while eliminating the need for developers to manually configure the fuzzers. This makes it a particularly appealing approach for non-experts who seek to improve the security of their OSS systems.




While continuous fuzzing offers some advantages over traditional methods, it is still unclear whether the effectiveness of shorter, more frequent fuzzing sessions is sustained across multiple repeated runs. It is also not obvious whether existing findings on traditional fuzzing can be generalized to continuous fuzzing or not. For example, traditional fuzzing involving long fuzzing sessions on a specific commit might allow a fuzzer to reach deeper parts of the code (\ie higher coverage), leading to more fuzzing bugs being found~\cite{cmc.2023.042361}. While prior work~\cite{DBLP:conf/sigsoft/BohmeF20} and OSS-Fuzz~\cite{OSSFuzzD2:online} have proven that short fuzzing sessions are also effective, it remains unclear how repeated short fuzzing sessions in continuous fuzzing can progressively increase code coverage and bug detection.


In this study, we aim to investigate and clarify how the number of fuzzing sessions can impact \defect detection and code coverage. To do so, we analyze historical fuzzing data from 878 open-source projects participating in the OSS-Fuzz initiative. To the best of our knowledge, this is the first large-scale study investigating the effectiveness of continuous fuzzing. The main contributions of this study are as follows:

\begin{enumerate}
    \item \textbf{Empirical Analysis of Continuous Fuzzing at Scale. }We conducted a large-scale empirical investigation on the effectiveness of continuous fuzzing by analyzing more than 3.95 million build logs, 1 million coverage reports, and over 70,000 issue reports.
    \item \textbf{Trends in Coverage and Fuzzing Bug Detection Rate Over Time. }We examined how the number of continuous fuzzing sessions relates to fuzzing bug detection rates and coverage trends, and found that many fuzzing bugs are detected in the early stages and that coverage generally increases over time.
    \item \textbf{Impact of Coverage Fluctuations on Fuzzing Bug Discovery. }We analyzed coverage changes at the time of fuzzing bug detection, and found that fuzzing bugs are more likely to be discovered when the coverage not only increased but also decreased.
    \item \textbf{Replication packages for Open Science.} To facilitate replication and further
studies, we provide the data used in our replication package.\footnote{
During peer review, the following limited public share link on Zenodo is provided: \url{https://bit.ly/3IQ8j0k}.}
\end{enumerate}


\smallskip
The remainder of this paper is organized as follows. Section \ref{sec:background} provides background information. Section \ref{sec:research_questions} states the research questions. Section \ref{sec:design} outlines the study design. Section \ref{sec:result} presents the results. Section \ref{sec:implication} discusses the implications. Section \ref{sec:validity} addresses threats to validity. Section \ref{sec:related} reviews related work. Section \ref{sec:Conclusion} concludes the paper.

\begin{figure*}[t]
    \centering
    \includegraphics[width=.9\linewidth]{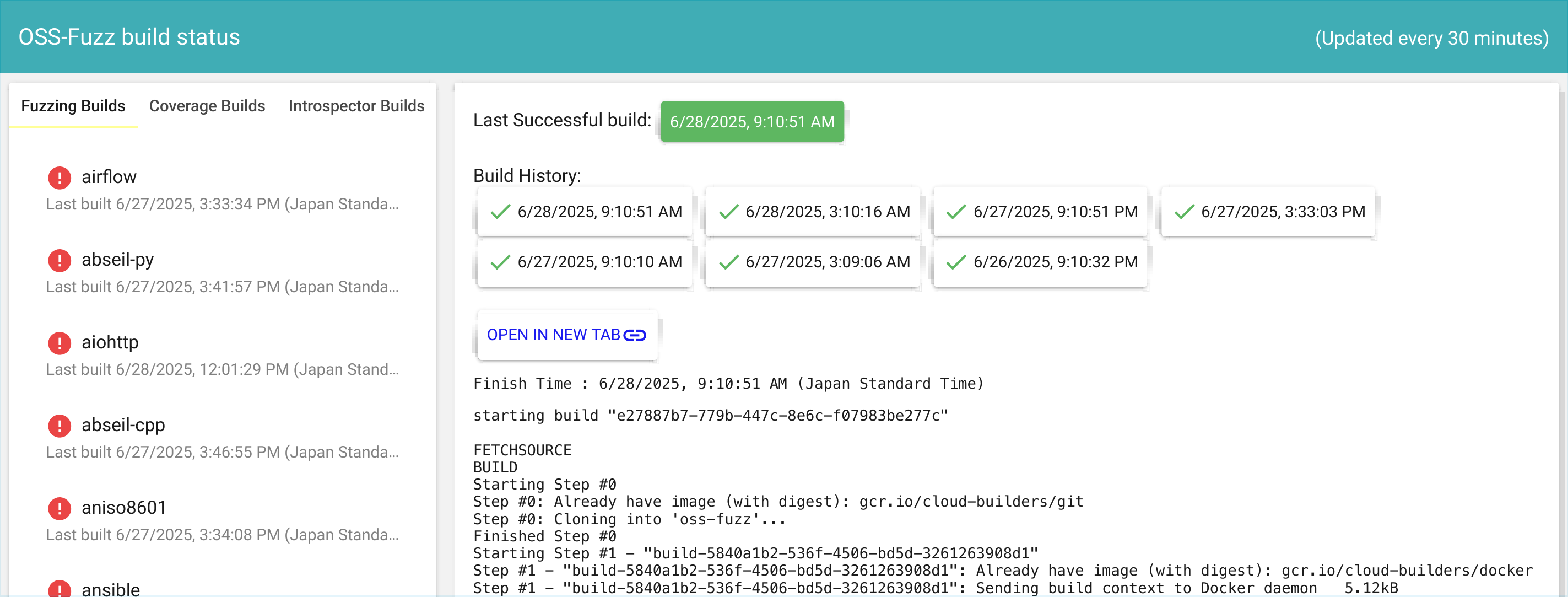}
    \caption{List of Build Logs on OSS-Fuzz}
    \label{fig:build_log}
\end{figure*}
\begin{figure}[t]
\captionsetup[lstlisting]{skip=0.4\baselineskip}
\begin{lstlisting}[numbers=left,caption={Example of Build Log},label={lst:buildlog_content}, escapechar=!]
Starting Step #2 - "srcmap"
Step #2 - "srcmap": Already have image: gcr.io/oss-fuzz/avahi
(ommited...)
Step #2 - "srcmap": {
Step #2 - "srcmap":   "/src/avahi": {
Step #2 - "srcmap":     "type": "git",
Step #2 - "srcmap":     "url": "https://github.com/lathiat/avahi",
Step #2 - "srcmap":     "rev": "fd482a74625b8db8547b8cfca3ee3d3c6c721423"
Step #2 - "srcmap":   },
Step #2 - "srcmap":   "/src/aflplusplus": {
Step #2 - "srcmap":     "type": "git",
Step #2 - "srcmap":     "url": "https://github.com/AFLplusplus/AFLplusplus.git",
Step #2 - "srcmap":     "rev": "c208dcf9c573e3d85990c7dea777646f7fa4961c"
Step #2 - "srcmap":   }
Step #2 - "srcmap": }
Finished Step #2 - "srcmap"

\end{lstlisting}
\end{figure}
\begin{figure*}[t]
    \centering
    \includegraphics[width=0.95\linewidth]{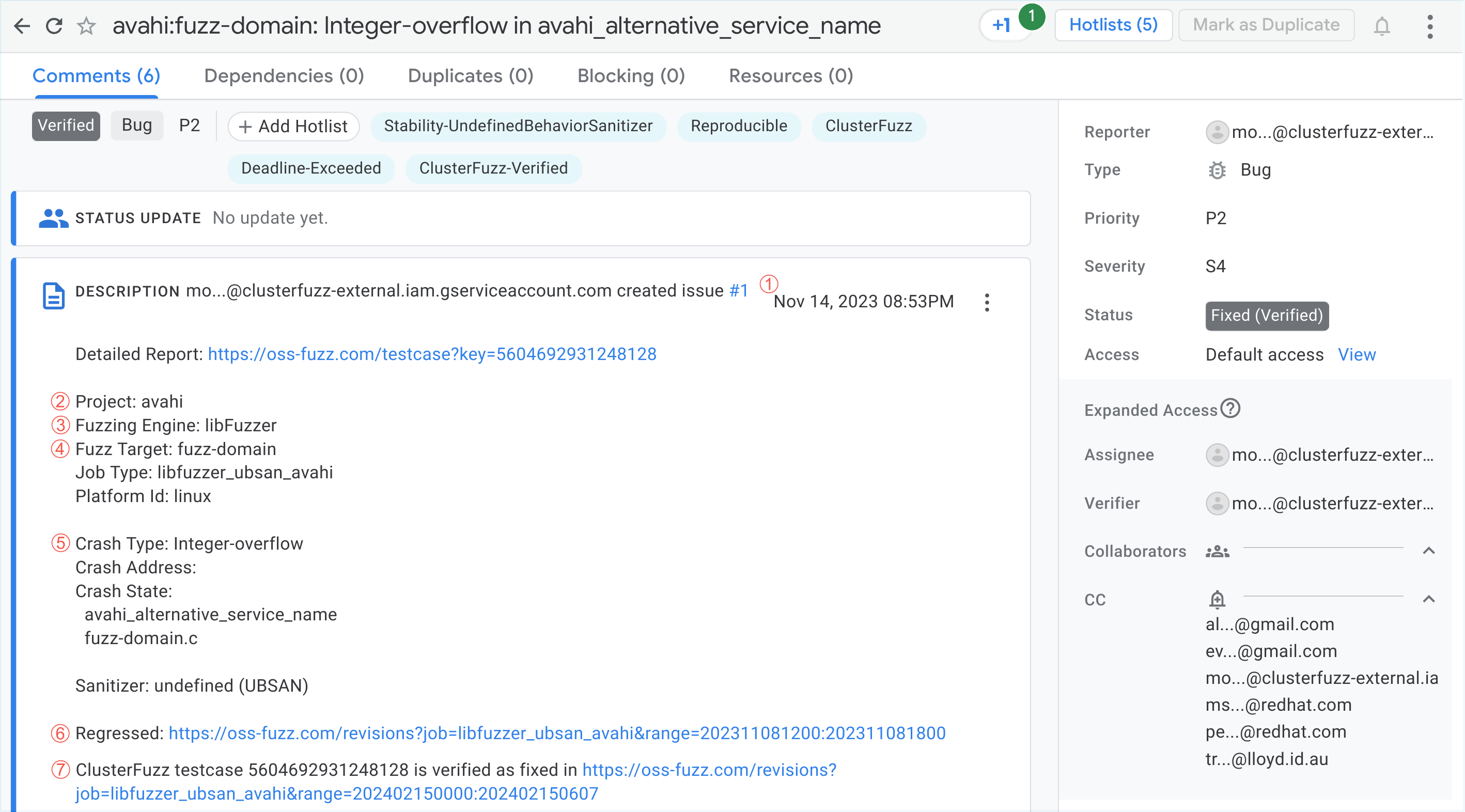}
    \caption{Example of Issue Report on the Issue Tracker for OSS-Fuzz}
    \label{fig:Issue_Tracker}
\end{figure*}

\section{Core Concept in OSS-Fuzz}\label{sec:background}
This section explains the concept of fuzzing and presents OSS-Fuzz, an implementation of continuous fuzzing. 

\subsection{Fuzzing}\label{sec:fuzzing}
Fuzzing, an automated software testing technique proposed by Miller\etal in 1990, is designed to trigger unintended behaviors intentionally~\cite{DBLP:journals/cacm/MillerFS90}. It generates a large volume of input data based on predefined rules, which is then fed into a program to verify its stability. By triggering unexpected crashes or security issues such as memory leaks, fuzzing is able to reveal security flaws in software systems. Despite its simple mechanism, fuzzing can uncover many vulnerabilities and bugs that other testing methods might miss.

Many studies have demonstrated the effectiveness of fuzzing at revealing fuzzing bugs. 
B{\"{o}}hme\etal\cite{DBLP:conf/icse/BohmeSM22} showed that as a fuzzer explores more of the codebase, the number of fuzzing bugs discovered tends to increase proportionally, highlighting a positive correlation between code coverage and bug detection. 
Liyanage\etal\cite{DBLP:conf/icse/LiyanageBTL23} conducted a seven-day fuzzing experiment and observed a consistent increase in code coverage over time. Their analysis indicated that code coverage growth tends to follow a logarithmic pattern. 

Despite its effectiveness, fuzzing has many pitfalls. Specifically, fuzzing is notorious for being resource-intensive, requiring substantial computational power and execution time~\cite{DBLP:journals/tosem/NourryKLBLK24}\cite{DBLP:conf/sigsoft/BohmeF20}. Previous studies recommend at least 24 hours of fuzzing to efficiently detect fuzzing bugs \cite{DBLP:conf/icse/BohmeSM22}\cite{DBLP:conf/ccs/KleesRCW018}. Additionally, Nourry\etal\cite{DBLP:journals/tosem/NourryKLBLK24} have shown that for fuzzing users, the cost and the complexity of setting up and preparing the environment pose significant challenges.


\subsection{Continuous Fuzzing and OSS-Fuzz}\label{sec:oss_fuzz}
Continuous fuzzing integrates fuzzing into the CI pipeline, ensuring that fuzzing sessions are conducted automatically and frequently, such as when new code is committed or modifications are made to the software. Unlike traditional fuzzing, which might be run periodically or on-demand, continuous fuzzing operates in short and frequent cycles, allowing for rapid identification of fuzzing bugs as soon as they are introduced into the codebase. Continuous fuzzing tools automatically execute fuzz tests, monitor the program's behavior, and report any anomalies or crashes. This automation helps maintain a high level of security throughout the development process. By continuously running fuzz tests, developers can detect and address fuzzing bugs quickly, thereby reducing the risk of security issues in the final product.

As of 2025, the most popular implementation of continuous fuzzing is the OSS-Fuzz service, which is developed and maintained by Google. OSS-Fuzz enables distributed and continuous fuzzing by leveraging ClusterFuzz, a platform to conduct fuzzing in a distributed manner. Additionally, OSS-Fuzz utilizes CIFuzz, a system designed to integrate fuzzing into CI/CD pipelines, allowing fuzzing to be performed automatically on a per-commit basis or at scheduled intervals. OSS-Fuzz has integrated over 1,000 open-source projects, contributing to the identification and remediation of more than 10,000 vulnerabilities and 36,000 bugs~\cite{OSSFuzzD2:online}.

Figure \ref{fig:oss_fuzz_process} illustrates the bug detection process using OSS-Fuzz.
First, open-source developers upload a build configuration of their software repository and the fuzz targets to OSS-Fuzz. Based on the build configuration, OSS-Fuzz periodically builds the target OSS project and the fuzz targets (Step 1. Pull \& Build in the figure). 

At the end of the build process, the compiled project files and fuzz targets are uploaded to cloud storage (Step 2. Upload product). The fuzzing build logs are also generated and stored in a Google Cloud Storage (GCS) bucket. Figure~\ref{fig:build_log} shows a build status dashboard for OSS Fuzz, and Snippet~\ref{lst:buildlog_content} shows an example of build logs that can be viewed in the dashboard.\footnote{\url{https://oss-fuzz-build-logs.storage.googleapis.com/log-004ed00b-bcd3-4c2e-bf70-9f9ff046cba8.txt}} 
The build log contains the project name (line 2 in Snippet~\ref{lst:buildlog_content}), required component names (\eg fuzzers and fuzz targets) (line 5, 10), Git commit SHA (line 8, 13),  error messages if any occurred, etc. 
Additionally, metadata files containing additional information such as build dates are stored in JSON files in GCS buckets.\footnote{\url{https://www.googleapis.com/storage/v1/b/oss-fuzz-gcb-logs/o/log-004ed00b-bcd3-4c2e-bf70-9f9ff046cba8.txt}} 

After the build is completed, the compiled files are uploaded to cloud storage. ClusterFuzz then downloads these files along with the initial fuzzing inputs and starts the fuzzing process (Step 3. Deploy) and fuzzing is applied to the uploaded fuzz targets (Step 4. Fuzz). 

Once ClusterFuzz triggers a bug, an issue report is automatically generated and submitted to the issue-tracking system (Step 5. Generate). Figure~\ref{fig:Issue_Tracker} shows an example of a issue report. An issue report contains the reported date (\ding{172}), 
project name (\ding{173}), 
fuzzing engine (\ding{174}), fuzz target (\ding{175}), 
and various crash information (\ding{176}) such as the type of crash. 
In addition, the reports contain the regression range (\ding{177}) 
and fixed range (\ding{178}), which shows the commit SHAs when the bug was detected and when the bug could not be reproduced. Note that issue reports remain private for 90 days after detection or until the bug is fixed to give time to developers to fix the bug and prevent any exploitation from malicious actors.

\begin{figure}[t]
    \centering
    \includegraphics[width=0.95\linewidth]{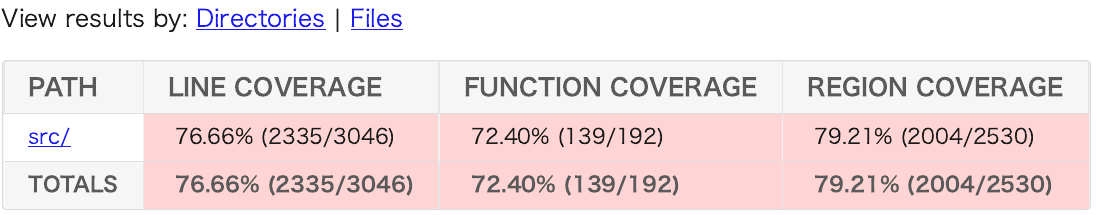}
    \caption{Example of Coverage Reports\protect\footnotemark}
    \label{fig:coverage_report}
\end{figure}
\footnotetext{\url{https://storage.googleapis.com/oss-fuzz-coverage/avahi/reports/20250526/linux/report.html}}
In addition, Fuzz Introspector analyzes the fuzzing results and generates daily coverage reports, which are also uploaded to the GCS bucket (Step 6. Analyze).\footnote{\url{https://storage.googleapis.com/oss-fuzz-introspector/avahi/inspector-report/20250526/fuzz_report.html}} The coverage reports include ``line coverage,'' ``function coverage,'' and ``region coverage.''\footnote{Note: OSS-Fuzz does not provide official information, but their source code suggests that region coverage is similar to branch coverage.}$^{,\thinspace}$\footnote{The types of coverage that can be measured by OSS-Fuzz vary depending on the programming languages of the fuzz target.}
Line coverage measures the proportion of source code lines executed during testing. Function coverage quantifies the number of functions that were invoked, while region coverage—closely related to branch coverage captures the extent to which distinct control-flow regions were exercised. For instance, the coverage report presented in Figure~\ref{fig:coverage_report} indicates that  76.66\% of the lines, 72.40\% of the functions, and 79.21\% of regions were covered during fuzz testing.





\section{Research Questions}\label{sec:research_questions}

This section describes the three research questions (RQs) we address in this study.

\smallskip
\noindent\textit{\rqA}\par
Continuous fuzzing, including OSS-Fuzz, performs fuzzing for a fixed duration each day, typically around 10 minutes by default \cite{ossfuzz-ci}. If the likelihood of detecting bugs continues to increase or remains consistently high as the number of fuzzing sessions accumulates, it would indicate that continuous fuzzing is effective over the long term and justifies sustained execution. In contrast, if most fuzzing bugs are detected during the initial sessions and the detection rate significantly declines thereafter, it would suggest that the majority of issues are uncovered early, and that fuzzing resources could be more efficiently allocated or strategically scheduled. However, the relationship between the cumulative number of fuzzing sessions and the bug detection rate remains unclear. RQ1 investigates how the number of detected fuzzing bugs is associated with the cumulative number of fuzzing sessions.

\smallskip
\noindent\textit{\rqB}\par
Coverage is a widely used performance metric for evaluating fuzzing techniques \cite{DBLP:journals/compsec/ChenCMWGL18}. Previous studies \cite{DBLP:conf/icse/LiyanageBTL23, DBLP:conf/sigsoft/BohmeMC20} have shown that, for traditional fuzzing, a single extended fuzzing session improves code coverage. On contrary, ClusterFuzz, which is used by OSS-Fuzz, includes a feature called Corpus Pruning~\cite{clusterfuzz_corpus_pruning}, which removes unnecessary inputs from the corpus while preserving coverage with a minimal set of test cases.
As more fuzzing sessions are performed, this pruning process helps refine the corpus, potentially allowing OSS-Fuzz to achieve progressively higher coverage over time. However, to the best of our knowledge, it remains unclear how continuous fuzzing's short, repetitive fuzzing sessions contribute to coverage improvement. 
If coverage increases consistently with the number of fuzzing sessions, it would suggest that even shorter time of fuzzing helps explore new program paths and justify sustained fuzzing efforts. 
RQ2 investigates how the number of fuzzing sessions impacts code coverage. 

\smallskip
\noindent\textit{\rqC}\par
OSS-Fuzz utilizes state-of-the-art fuzzers such as AFL and libfuzzer, both of which use coverage-guided fuzzing (\ie they aim to maximize code coverage to enhance fuzzing bug detection). However, despite the central role of coverage in modern fuzzing strategies, the relationship between coverage and the bug detection rate remains unclear. Since OSS-Fuzz provides the fuzzing results for each short session, including the changes in coverage and found fuzzing bugs, we are able to investigate how an increase or decrease in coverage correlates with more or fewer fuzzing bugs detected. RQ3 analyzes the changes in coverage when fuzzing bugs are triggered to see if increasing, decreasing, or maintaining coverage affects bug discovery.

\section{Study Design}\label{sec:design}
This section describes the approaches for the data collection, project selection, and data analyses to answer each RQ.

\subsection{Data Collection}\label{sec:data_collection}
We first data mined all issue reports from the official OSS-Fuzz issue tracker.
This process gave us a total of \gotReportNumber reports spanning over \gotReportProject projects. We then excluded all issue reports that did not have a ``Fixed" status (\ie unfixable, open, unreproducible, etc.), which resulted in \FixedReportNumber valid reports from \FixedReportProject projects. 
Note that issues in the tracker are classified into two categories: Bugs and Vulnerabilities. In this study, we consider both categories and collectively refer to them as ``fuzzing bugs'' for simplicity, following the previous studies~\cite{DBLP:conf/msr/KellerMM23}.

Next, we collected over \totalBuildNumber build logs generated from compiling the fuzzers and the projects between \buildStartDate and \buildEndDate in order to measure how many fuzzing sessions had been performed. 
Finally, we then collected all the coverage reports generated by \texttt{Fuzz-Introspector} and obtained a total of \totalCoverageNumber coverage reports.  We extracted the line coverage information from the coverage reports.

\subsection{Project Selection}\label{sec:project_collection}
As of \ossFuzzProjectDate, \ossFuzzProjectNumber open-source projects have been fuzzed by OSS-Fuzz. Since we are investigating the evolution of code coverage and fuzzing bugs found over time, we had to select projects with sufficient historical data to conduct our analyses. By default, OSS-Fuzz will fuzz each participating project at least once a day but project maintainers can change the default configuration to be fuzzed more frequently. 
To ensure we are using projects with sufficient historical data, we therefore selected projects that had measured their coverage at least 365 times  by \targetProjectDate (meaning they had been fuzzed for at least 1 year by OSS-Fuzz). This filtering resulted in \targetProjectNumber valid projects, accounting for \targetProjectRate of all projects that have participated in OSS-Fuzz. By narrowing down the issue reports to those associated with these projects, we were eventually able to analyze \ProjectReportNumber issue reports from \ProjectReportProject projects.

\subsection{Linking issue reports and coverage reports}\label{sec:data_linking}
\begin{figure}[t]
    \centering
    \includegraphics[width=0.95\linewidth]{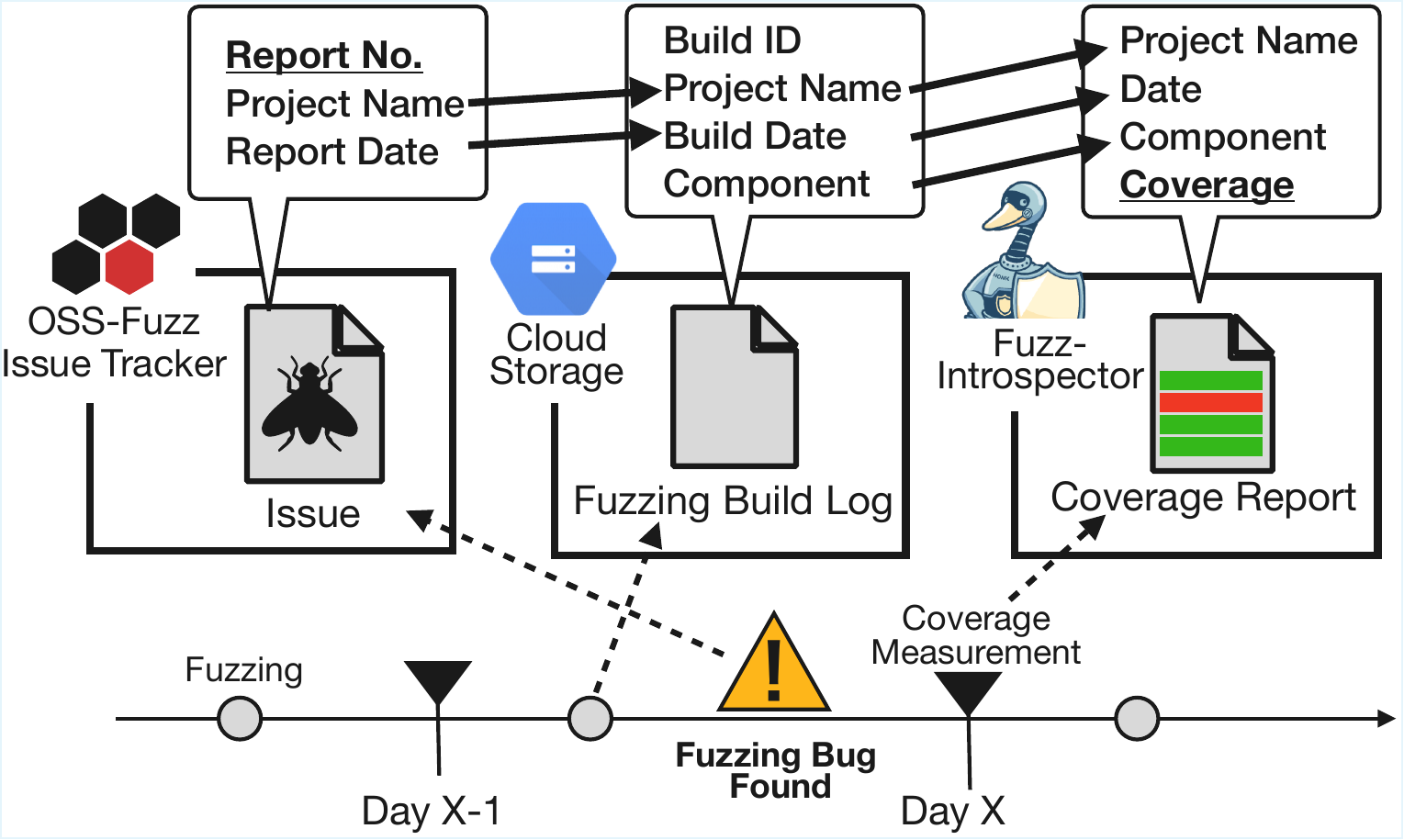}
    \caption{Overview of Data Linking Process}
    \label{fig:overview_data_acquisition}
\end{figure}

In RQ3, we aim to link issue reports with coverage reports. However, issue reports lack the necessary data for this linkage. To address this, we utilize fuzzing build logs. Figure \ref{fig:overview_data_acquisition} provides an overview of the data linking process.

We first associate each issue report with its corresponding build log using the project name and the date specified in the issue reports. Since issue reports are automatically generated by OSS-Fuzz when a fuzzer crashes the fuzz target, we can identify the corresponding build log by selecting the last successful build log that was generated before the report was created. Using this approach, we successfully associated \connectVulnsAndBuilds  reports with their corresponding build logs (\connectVulnsAndBuildsPercent\%). The remaining \cantConnectVulnsAndBuildsPercent\% could not be associated due to build failures, etc. 

Next, we link fuzzing build logs with coverage reports by extracting properties from the build logs. Specifically, we use the project name, build timestamp, component path, and component revision (\ie SHA) found in the build logs to match each log with its corresponding coverage report.

However, in several cases, changes are made to the component (\ie the project) between the time when fuzzing is conducted and the time when the coverage report is generated. These changes cause the project's SHA to be updated and therefore causes a mismatch between the build log's identifier and the coverage report's identifier. For these cases, even if all other information matches (\eg name, timestamp, component path, etc.) we cannot guarantee that both the build log and the coverage report were executed on the same revision of the project which affects the validity of our results. To ensure the accuracy of our results, we only include issue reports where the issue report's SHA identifier has both a build log and a coverage report with a matching identifier. This additional filtering process resulted in \connectVulnsAndCoverageNumber valid issue reports for RQ3.

\subsection{Data Analysis}\label{sec:analysis_method}
This section describes the data analysis process to answer each RQ.
\smallskip

\subsubsection{\textbf{RQ1. Fuzzing Sessions and Fuzzing Bug Detection Rate}}
RQ1 investigates the relationship between the number of cumulative fuzzing sessions and the fuzzing bug detection rate. Specifically, we calculate the percentage of projects that detect fuzzing bugs during each fuzzing session (\eg the first, second, $i$-th session). To achieve this, we examine the build logs of each project at session $i$
to identify whether fuzzing bugs are detected during the $i$-th fuzzing session. We then calculate the percentage of projects that detect fuzzing bugs, referred to as the bug detection rate $R_i$.
The bug detection rate  $R_i$ at a given $i$-th fuzzing session is formulated as follows:

\begin{equation}\label{eq:project_rate}
    R_i = \frac{P_i}{F_i},
\end{equation}
where $P_i$ represents the number of projects that detected a fuzzing bug during the $i$-th fuzzing session, and $F_i$ represents the number of projects that performed fuzzing at the $i$-th session. To maintain a sample size of at least 100 projects for our analysis, we calculated the rate of bug detection up to the \RQoneMaxFuzzing session.

\medskip
\subsubsection{\textbf{RQ2. Fuzzing Sessions and Coverage}}
RQ2 investigates how code coverage evolves as the number of cumulative fuzzing sessions increases. Specifically, we aim to understand both the general and project-specific trends in coverage growth over time.

\begin{itemize}
    \item \textit{Overall trend:} We examine the overall relationship between the number of fuzzing sessions and line coverage across all studied projects. Specifically, we analyze the line coverage scores at each cumulative fuzzing session and assess whether the median coverage exhibits a consistent trend as the number of fuzzing session increases.
    \item \textit{Project specific trend:} We further investigate the correlation between the number of fuzzing sessions and line coverage on a per-project basis. For each project, we compute the correlation coefficient and then analyze the distribution of these coefficients across all projects, reporting the median as a summary statistic. 
    
\end{itemize}

To quantify the relationship between fuzzing sessions and code coverage, we employ the Spearman’s rank correlation coefficient, which is a nonparametric measure to account for potential non-linear trends. To ensure statistical robustness, we exclude sessions for which coverage data is available from fewer than 100 projects. Our analysis therefore includes data up to the \RQtwoMaxCoverage fuzzing session.

\medskip
\subsubsection{\textbf{RQ3. Coverage Changes and Fuzzing Bug Detection}}
RQ3 investigates the impact of the coverage changes on the fuzzing bug detection. We first identify fuzzing sessions where fuzzing bugs were detected by identifying whether the sessions are associated with issue reports or not (\ie Section~\ref{sec:data_linking}). We then obtained the coverage values of each session and calculated the difference in coverage before and after fuzzing.

Next, we conducted a statistical analysis to examine whether the distributions of coverage changes differ significantly between sessions that detected a bug and those that did not. We employed the Brunner-Munzel test~\cite{BrunnerTest}, which is a non-parametric statistical test, to detect the statistically significant difference between the two groups. The reason why we use this test is that we found that the variances between the two groups were not equal\footnote{We confirmed this with Levene’s test ($p < 0.01$)~\cite{LEVENE_TEST}.}, violating the homoscedasticity assumption required by the Mann-Whitney U test~\cite{KASUYA20011247}. On the contrary, the Brunner-Munzel test does not assume equal variances but can assess the significance of differences between groups, as the typical statistical tests such as Mann-Whitney U test do.


\pdfcompresslevel=9
\pdfobjcompresslevel=3

\section{Result}\label{sec:result}
In this section, we present the results for each RQ.

\begin{figure}[t]
    \centering
    \includegraphics[width=.96\linewidth]{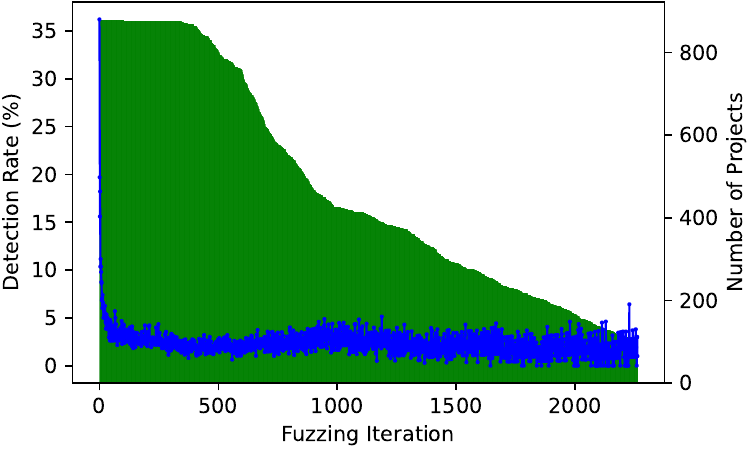}
    \caption{Bug Detection Rate Over Fuzzing Sessions}
    \label{fig:rq2_project}
\end{figure}

\subsection*{\rqA}
Figure \ref{fig:rq2_project} shows both the proportion ($y$-axis left side) and absolute ($y$-axis right side) number of projects in which fuzzing bugs were detected, with the corresponding cumulative number of fuzzing sessions at the time of detection. The green shaded graph indicates the absolute number of projects that reached the number of fuzzing sessions shown on the $x$-axis. The blue line graph represents the proportion of projects (left side $y$-axis) that detected a fuzzing bug during the $i$-th session, shown on the $x$-axis. 

Our results indicate that fuzzing bugs are more frequently detected during the early stages (\ie early fuzzing sessions). Notably, for almost \RQoneFirstDetectionRate of our studied projects, a fuzzing bug was discovered during the first fuzzing session after the project was integrated into OSS-Fuzz. Although this detection rate drops to \RQoneSecondDetectionRate in the second session, subsequent fuzzing sessions also often find fuzzing bugs. These results show that a significant portion of open-source software projects contain fuzzing bugs and could benefit from adopting fuzz testing as part of their testing practices.

Our results also indicate that during the \RQoneDecreaseCount fuzzing session, the detection rate drops to \RQoneDecreaseDetectionRate\%, falling below the 5\% threshold for the first time. Subsequently, it remains relatively stable, exhibiting a slight downward trend over time. After the \RQoneDecreaseCount session, the detection rate stabilizes around a median of \RQoneQmedian, with most of the remaining data fluctuating close to the median value.
These results demonstrate that although the efficiency of fuzzing declines from early to later sessions, fuzzers remain effective at discovering new fuzzing bugs even after prolonged operation and thousands of fuzzing sessions.

\begin{dbox}
\rqa . \it 
Fuzzing bugs are most frequently detected during the first few fuzzing sessions, with \RQoneFirstDetectionRate of projects detecting a fuzzing bug during the first session and \RQoneSecondDetectionRate during the second. Detection rates drop below 5\% by the \RQoneDecreaseCount session, then remain consistent around the median \RQoneQmedian detection rate.
\end{dbox}

\begin{figure}[t]
    \centering
    \includegraphics[width=.96\linewidth]{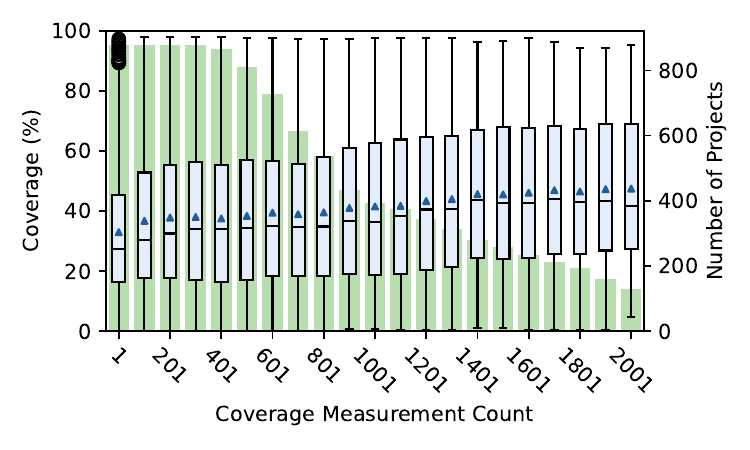}
    \caption{Cumulative Fuzzing Sessions and Coverage}
    \label{fig:rq1_box}
\end{figure}

\subsection*{\rqB}
Figure \ref{fig:rq1_box} depicts the distribution of coverage levels based on the cumulative number of fuzzing sessions. The median line denotes the median coverage percentage and the triangle shows the mean coverage percentage. Additionally, the bar graphs represent the number of projects that were analyzed for the $i$-th fuzzing session shown on the $x$-axis.

We observed the lowest median coverage values during the initial fuzzing sessions. Notably, in the early phase of continuous fuzzing (\ie sessions \RQtwoInitialPhase), although the overall coverage remained low, the median coverage exhibited the steepest increase, indicating the highest rate of coverage growth. After this period, although the growth rate decreased, a stable and consistent increase in coverage was observed. Our observations align with the findings of Liyanage\etal\cite{DBLP:conf/icse/LiyanageBTL23}, where they found that traditional fuzzing showed no signs of reaching an asymptotic upper bound for coverage. 
Interestingly, while traditional fuzzing tends to exhibit logarithmic growth in coverage over time, our results on continuous fuzzing seem to display a linear increase in coverage as the number of fuzzing sessions increases.

We examined the relationship between the number of fuzzing sessions and median code coverage across all projects by computing Spearman’s rank correlation coefficient. The resulting coefficient was \RQtwoCorr, indicating a very strong positive correlation between the two metrics. This suggests that, in general, continued fuzzing tends to improve coverage.

\begin{figure}[t]
    \centering
    \includegraphics[width=.96\linewidth]{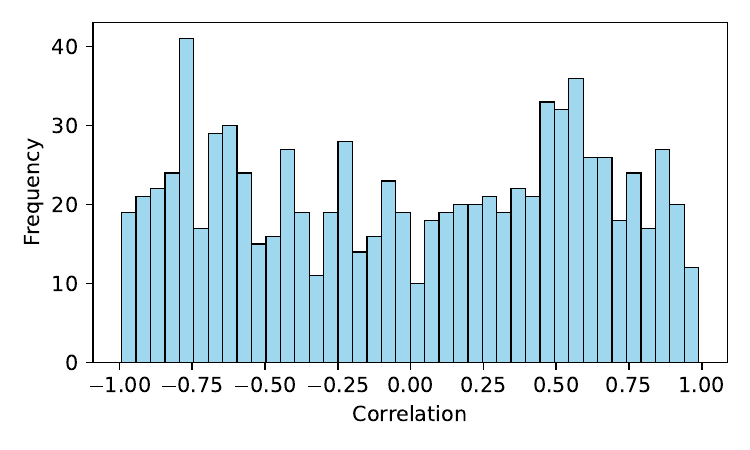}
    \caption{Correlation Histogram for Each Project}
    \label{fig:rq2_hist}
\end{figure}

However, a different pattern emerges when we compute the correlation within individual projects. Figure \ref{fig:rq2_hist} shows the distribution of correlation coefficients across the studied projects. These coefficients range from strongly negative to strongly positive, with a median value of \RQtwoMedianCorrEachProject. 
\begin{figure}[t]
    \centering
    \includegraphics[width=.96\linewidth]{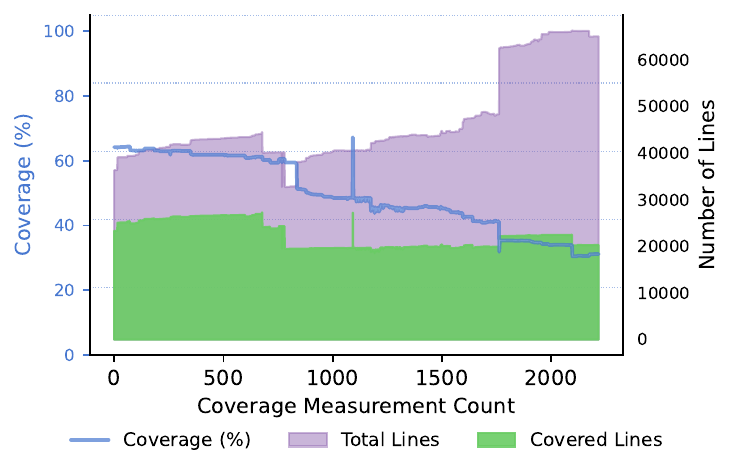}
    \caption{Decreasing coverage with the covered lines and total lines of code. }
    \label{fig:rq2_exam_projct}
\end{figure}

One plausible explanation for negative correlations in some projects is that the total number of lines of code increases over time due to ongoing development, while the number of lines executed during fuzzing remains relatively stable. As a result, the relative coverage decreases, leading to a negative correlation. For instance, in project \texttt{\RQtwoExampleProjectName}, we analyzed \RQtwoExampleProjectCount coverage reports and observed a correlation coefficient of \RQtwoExampleProjectCorr. As illustrated in Figure \ref{fig:rq2_exam_projct}, although the total codebase size steadily increased (\ie purple shade), the number of executed lines remained nearly constant (\ie green shade), resulting in a gradual decline in coverage (\ie blue line). 

\begin{dbox}
\rqb . \it 
Continuous fuzzing steadily increases code coverage, with the most rapid gains occurring early on. While the number of fuzzing iterations and coverage show a strong positive correlation overall, project-level analyses reveal substantial variability influenced by factors like codebase growth.
\end{dbox}

\subsection*{\rqC}

\begin{figure}[t]
  \centering
  \begin{subfigure}{0.48\linewidth}
    \centering
    \includegraphics[width=0.7\linewidth]{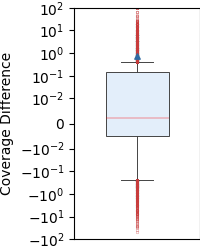} 
    \caption{Fuzzing sessions that detected fuzzing bugs}
    \label{fig:rq3_box_detect}
  \end{subfigure}
  \hfill
  \begin{subfigure}{0.48\linewidth}
    \centering
    \includegraphics[width=0.7\linewidth]{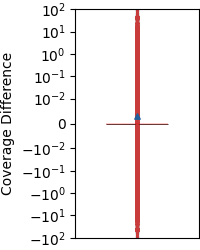} 

    \caption{Fuzzing sessions that did not detect fuzzing bugs}
    \label{fig:rq3_box_not_detect}
  \end{subfigure}
  \caption{Distribution of Coverage Changes}
  \label{fig:rq3_box}
\end{figure}

Figure \ref{fig:rq3_box} presents the distribution of coverage changes associated with the detection of fuzzing bugs. Our analysis reveals that the median coverage change is 0\%, regardless of whether fuzzing bugs were detected. However, the interquartile ranges (IQRs) of the two distributions differ notably. When fuzzing bugs are identified, the IQR spans from \RQthreeDetectedIRQsFirst to \RQthreeDetectedIRQsEnd, indicating a broader spread of coverage changes. In contrast, when no fuzzing bugs are found, the changes are tightly clustered around 0.00\%, suggesting minimal variation. While these percentage differences may appear small, they are substantial in the context of large codebases (median size: \RQthreeTotalLine lines of code). 
We applied the Brunner-Munzel test and observed a statistically significant difference between the two groups ($W = -19.64, p < 0.01$).



\begin{table}[bht]
    \small
    \centering
    \caption{Percentages of sessions that experienced an increase, decrease or stagnation of coverage when at least one fuzzing bug is found/not found in a session}
    \begin{tabular}{crr}
        \toprule
                 & \multicolumn{1}{c}{Found} & \multicolumn{1}{c}{Not Found} \\ \midrule
        Increase & \textbf{51.6\%}    & 20.6\%        \\ 
        Stagnation   & 20.5\%    & \textbf{61.0\%}        \\ 
        Decrease & 27.9\%    & 18.4\%        \\ \bottomrule
    \end{tabular}
    \label{tab:coverage_changes}
\end{table}


Additionally, we classified the coverage changes into three categories: \texttt{Increase}, \texttt{Stagnation}, and \texttt{Decrease}. 
Specifically, we define \texttt{Stagnation} as no change in coverage, \texttt{Increase} as at least one additional line of code being covered, and \texttt{Decrease} as at least one previously covered line no longer being covered. 
Table~\ref{tab:coverage_changes} summarizes the frequency of each event. Our results show that when fuzzing bugs were found, approximately 52\% of fuzzing sessions exhibited increased coverage, 28\% showed decreased coverage, and 21\% experienced stagnation. In contrast, when no fuzzing bugs were found, only 21\% of sessions showed increased coverage, while 61\% remained stable and 18\% showed decreased coverage. These findings suggest that both coverage increases and decreases are associated with bug detection, highlighting that not only coverage growth but also reduction may signal meaningful exploration behavior.


\begin{dbox}
\rqc . \it 
When fuzzing bugs are discovered, coverage increases in about 52\% of cases, decreases in about 28\% of cases, and remains stagnant in the remaining 21\% of cases.
\end{dbox}

\section{Implication}\label{sec:implication}
Our empirical findings can help improve continuous fuzzing platforms and provide insights for practitioners and future research. We present the following key implications.

\subsection{Implication for continuous fuzzing providers (platforms)}
Current continuous fuzzing platforms allow their users to set up a static duration of fuzzing. However, these platforms lack adaptive mechanisms that would allow adjusting the fuzzing duration and frequency based on a project's testing history or coverage achieved in previous runs. Findings from RQ1 indicate that fuzzing bugs are more likely to be discovered during the early stages of fuzzing adoption, with diminishing returns over time. This suggests that fuzzing duration should be adaptively shortened or extended based on the cumulative number of sessions. Additionally, RQ3 shows that fuzzing bugs are often discovered not only when coverage increases but also when it decreases, such as after the integration of large code changes. This implies that fuzzing duration should also scale with the magnitude of recent codebase modifications.
Implementing such dynamic adjustments would enhance both the effectiveness and efficiency of fuzzing, while also contributing to energy savings. Therefore, we recommend that \textit{continuous fuzzing providers develop functionality to dynamically adjust fuzzing duration based on cumulative fuzzing history and the size of code changes.}

\subsection{Implication for Practitioners}
Our analysis reveals that \RQoneFirstDetectionRate of OSS-Fuzz projects detected at least one fuzzing bug during the first fuzzing session, and \RQoneSecondDetectionRate during the second, indicating that many fuzzing bugs exist prior to the adoption of fuzzing. These findings highlight the value of early integration of fuzzing into the development lifecycle to enable rapid bug detection and mitigation, thereby reducing security risks and associated costs. We thus first recommend that \textit{practitioners should integrate fuzzing in their project's development process as early as possible. }

Furthermore, the bug detection rate remains above 5\% through the first 25 sessions, suggesting that \textit{practitioners should allocate additional resources during this early phase and after integrating significant code changes}. Even beyond the 25th session, the detection rate rarely drops to zero, highlighting the continued value of sustained, low-cost fuzzing efforts as coverage expands over time.




\subsection{Implication for Researchers}
RQ2 indicates that code coverage tends to increase linearly with prolonged fuzzing. Moreover, the likelihood of discovering fuzzing bugs rises when coverage changes. While increased coverage typically results from exploring new code regions, it is notable that fuzzing bugs were also uncovered during periods of coverage decline. Such declines are likely attributable to substantial changes in the codebase. This suggests that most existing fuzzing techniques are not optimized for continuous fuzzing environments, as they often fail to generate inputs that exercise newly added code. Therefore, \textit{researchers should propose more diff-aware fuzzing strategies that effectively target and test modified or newly introduced code regions in evolving software systems}.

\section{Threats to Validity}\label{sec:validity}
We identify several potential threats to the validity of our findings and discuss how they may impact the interpretation of our results.

\smallskip
\textbf{Internal Validity:}
Internal validity refers to the threats to validity concerning whether the observed outcomes in this study can be causally attributed to the variables under investigation. 
Open-source contributors may conduct local fuzzing sessions to identify and resolve fuzzing bugs before committing code to repositories. This pre-commit testing could reduce the number of fuzzing bugs subsequently detected through continuous fuzzing, potentially affecting our measurements.
In addition the projects examined in this study are predominantly high-profile OSS-Fuzz participants that likely undergo rigorous testing and code review processes within structured development environments. These mature development practices may result in lower baseline bug rates compared to typical open-source projects, potentially limiting the observed effectiveness of continuous fuzzing.

\smallskip
\textbf{Construct Validity:}
Construct validity refers to the threats to validity concerning whether the measurement methods used in this study accurately reflect the intended theoretical concepts or constructs. 
OSS-Fuzz records coverage data once daily, while our analysis relies on coverage measurements immediately before and after bug detection. When multiple fuzzing executions occur within a 24-hour period, the actual coverage at bug detection may differ from the recorded daily values. Although approximately 98\% of projects perform fuzzing only once daily, minimizing this impact, the temporal mismatch cannot be entirely eliminated.

Additionally, our analysis treats all detected fuzzing bugs equally, regardless of severity or exploitability. This approach may not accurately capture the security impact or reflect the relative effectiveness of continuous fuzzing for identifying critical versus minor fuzzing bugs.

\smallskip
\textbf{External Validity:}
External validity refers to the threats to validity related to the generalizability of our findings to other contexts or software projects. 
This study analyzed 878 projects. While this is a relatively large dataset, projects participating to OSS-Fuzz are core OSS projects, which implies that they have higher security requirements than average projects and therefore are more incentivized to conduct continuous fuzzing. As a result, conducting the same experiment on other projects might yield different results.

Additionally, our results may be specific to the particular fuzzing engines and strategies employed by OSS-Fuzz and may not generalize to other fuzzing approaches, tools, or methodologies used in different continuous fuzzing implementations. The development practices, contributor expertise levels, and security awareness of OSS-Fuzz participating projects may differ substantially from typical open-source projects or proprietary software development environments, potentially affecting the external validity of our conclusions.

\section{Related Work}\label{sec:related}
This section reviews prior empirical studies on both traditional and continuous fuzzing techniques.

\subsection{Empirical Studies on Traditional Fuzzing}
Several studies have characterized the fundamental patterns of coverage growth in traditional fuzzing.
B{\"{o}}hme\etal\cite{DBLP:conf/sigsoft/BohmeMC20} investigated the trend of coverage growth using FST, a benchmark suite for evaluating fuzzer performance. They demonstrated that for many projects, coverage increases sharply at the beginning of fuzzing and subsequently follows a logarithmic growth pattern. 
Liyanage\etal\cite{DBLP:conf/icse/LiyanageBTL23} conducted an experiment to observe coverage trends when fuzzing is executed for extended periods. Their results indicated that when fuzzing a single dataset over a long duration, coverage growth follows a logarithmic pattern.
Additionally, B{\"{o}}hme\etal\cite{DBLP:conf/icse/BohmeSM22} analyzed the correlation between coverage achieved by fuzzing tools and the number of detected bugs. They revealed a strong positive correlation between coverage and bug detection.

Building on understanding coverage patterns, researchers have developed methods to predict and estimate fuzzing outcomes. Liyanage\etal\cite{DBLP:conf/icse/LiyanageBTL23} proposed a dynamic estimation method for reachable code in fuzzing by applying species diversity estimation techniques. Their findings revealed that existing static analysis methods tend to either overestimate or underestimate the amount of reachable code. Liyanage\etal\cite{DBLP:conf/icse/Liyanage0TB24} proposed a method for predicting future coverage discovery rates in greybox fuzzing while accounting for adaptive bias. Their approach achieved up to an order of magnitude lower prediction error compared to existing methods. Furthermore, in predicting the time to reach a target coverage rate, their method demonstrated an average error reduction of 35–77\%. Böhme\etal\cite{DBLP:conf/sigsoft/BohmeLW21} proposed a method to estimate the residual risk of undiscovered bugs in greybox fuzzing campaigns and demonstrated that existing approaches significantly underestimate this risk. Their evaluation showed that the proposed method provides a more realistic estimation. 

Multiple approaches have been developed to improve fuzzing effectiveness and efficiency. Klees\etal\cite{DBLP:conf/ccs/KleesRCW018} compared the performance of multiple fuzzers across different execution time intervals. Their findings suggest that running fuzzing for at least 24 hours is optimal for maximizing bug detection efficiency. Zhou\etal\cite{DBLP:conf/kbse/ZhouWLLJ20} addressed the inefficiency of fuzzing by identifying the high cost of coverage tracing as a major bottleneck. They proposed a framework named Zeror, which balances speed and accuracy through dynamic binary switching. As a result, Zeror achieves an average 159.8\% speedup over AFL while preserving coverage information. 
Bundt\etal\cite{DBLP:conf/icst/BundtFDRL23} focused on human intervention in fuzzing and proposed a method that leverages compartment analysis to examine intermediate fuzzing results and identify concrete intervention points for improving coverage. Their approach demonstrated coverage improvements of up to 94\%, with a median increase of 13\%. 
Feng\etal\cite{DBLP:conf/icst/FengRWLKKS23} proposed MagicMirror, a fuzzing approach for smart contracts that leverages precondition-based input generation and selective state exploration, achieving an average of 21\% improvement in code coverage compared to existing methods.

The reliability of coverage measurement itself has also received attention. 
Liu\etal\cite{DBLP:conf/kbse/LiuWSYJ21} investigated the accuracy of instrumentation used by fuzzers and revealed that instrumentation errors are a common issue in real-world coverage-based greybox fuzzers. They further proposed InstruGuard, a technique for detecting and repairing such errors, and demonstrated a high repair success rate, averaging over 99.9\%. Moreover, their evaluation showed that the repaired instrumentation led to an increase in the number of executed paths and detected bugs during fuzzing.
Gao\etal\cite{DBLP:conf/fuzzing/GaoPLCMR23} conducted a systematic classification and analysis of fuzz blockers to investigate the root causes of the ``plateau" phenomenon in greybox fuzzing, where code coverage growth stagnates. Their study identified five categories of fuzz blockers and revealed that 61.7\% of the blockers are caused by the design of the fuzz driver, which is responsible for delivering inputs to the target program.

While prior studies have primarily examined the effectiveness of long-running fuzzing on specific versions of software projects, we investigate the impact of short, iterative fuzzing sessions conducted continuously throughout the software development lifecycle. We present the first large-scale empirical study that characterizes longitudinal trends in \defect detection rates and coverage evolution across thousands of cumulative fuzzing sessions on evolving software systems.

\subsection{Empirical Studies on Continuous Fuzzing}

Several studies have examined the characteristics of bugs and vulnerabilities detected through continuous fuzzing systems. 
Ding\etal\cite{DBLP:conf/msr/DingG21} analyzed bugs detected by OSS-Fuzz, specifically examining the types of detected bugs, the time required for detection, and the time taken for fixes.
Zhu\etal\cite{DBLP:conf/ccs/ZhuB21} investigated the characteristics of bugs detected by OSS-Fuzz. They found that 77\% of bugs detected by OSS-Fuzz were recently introduced, indicating that continuous fuzzing is particularly effective at identifying newly introduced security issues.
Keller\etal\cite{DBLP:conf/msr/KellerMM23} conducted a large-scale analysis of bug reports from OSS-Fuzz, examining the lifespan of bugs, the developers responsible for fixing them, and bug types. They recommended strategies to help developers learn more effectively while debugging code.

Research has also focused on optimizing fuzzing parameters and understanding infrastructure limitations. 
Klooster\etal\cite{DBLP:conf/icse/KloosterTBHB23} studied the relationship between fuzzing duration and bug detection rate, recommending 30 minutes as an optimal duration for continuous fuzzing.
Nourry\etal\cite{DBLP:journals/tosem/NourryKSSK25} investigated the issue of build failures in fuzzing. Their study revealed that 9.41\% of OSS-Fuzz project builds fail, highlighting challenges in maintaining a stable fuzzing infrastructure.

Recent work has specifically targeted fuzzing optimization within continuous integration and deployment environments. Sharma\etal\cite{DBLP:conf/fuzzing/SharmaCM24} proposed a method called PaZZER, which enables effective fuzzing in CI/CD environments by incrementally computing the estimated distance to the modified code regions based on the call graph. Compared to the conventional method AFLGo, PaZZER reduces the distance calculation time by up to 140 seconds and demonstrates effectiveness even in short tests of around 10 minutes for large-scale software systems that previously faced challenges in terms of processing time.
Zhang\etal\cite{DBLP:journals/iet-sen/ZhangCCYZL23} proposed a method called CIDFuzz, which focuses on code changes introduced during Continuous Integration (CI) and their impact range to achieve efficient fuzzing. By appropriately allocating testing resources based on control-flow and data-flow information derived through static analysis, CIDFuzz reduces the time required to achieve coverage of the modified code by up to 34.78\% compared to conventional methods.

In the domain of kernel fuzzing, Ruohonen\etal\cite{DBLP:conf/issre/RuohonenR19} analyzed the effectiveness of continuous kernel fuzzing using syzkaller/syzbot on Linux and BSD-based kernels. They reported that the median time-to-fix (TTF) for the Linux kernel was 38 days, while all BSD kernels had TTFs of less than 20 days. Also, only about 23\% of the fixed bugs in the Linux kernel underwent code review or additional testing.

While prior work has examined the detected bugs during continuous fuzzing or specific aspects of continuous fuzzing such as determining optimal session duration, our study emphasizes the effectiveness that emerges from the cumulative and sustained execution of fuzzing over time. We investigate how the total number of fuzzing sessions correlates with both \defect detection and coverage evolution. Additionally, we analyze the relationship between fluctuations in code coverage and the discovery of \defects. This analysis contributes to a deeper understanding of continuous fuzzing and offers actionable insights for enhancing its practical effectiveness.

\section{Conclusion}\label{sec:Conclusion}

This study presents the first large-scale empirical investigation on the effectiveness of continuous fuzzing, examining over 1 million fuzzing sessions from 878 open-source projects participating in OSS-Fuzz. Through our comprehensive investigation of \defect detection patterns, coverage evolution, and the relationship between coverage changes and \defect discovery, we provide several key new insights into continuous fuzzing behavior.

We found that \defects are predominantly detected during the initial phases of fuzzing adoption, with \RQoneFirstDetectionRate of projects detecting \defects in their first session and \RQoneSecondDetectionRate in their second session. This high early detection rate indicates that many \defects exist in codebases prior to fuzzing integration, emphasizing the importance of early adoption. While detection rates stabilize around 5\% after the \RQoneDecreaseCount session, continuous fuzzing remains effective even after thousands of sessions. 

Additionally, our analysis reveals that continuous fuzzing exhibits linear coverage growth over time, contrasting with the logarithmic patterns observed in traditional fuzzing. We found an overall correlation of \RQtwoCorr between the cumulative number of fuzzing sessions and coverage percentage, though some projects vary due to factors such as development activity and codebase growth.
Examining the relationship between coverage changes and \defect detection, we found that \defects are discovered when coverage increases (51.6\% of cases), decreases (27.9\%), or remains stable (20.5\%). This suggests that both coverage expansion and reduction signal meaningful code exploration leading to \defect discovery.

These findings have significant implications for practitioners, CF platforms, and researchers. The high rate of detection during early-stages of fuzzing suggests opportunities for adaptive fuzzing strategies, while the linear coverage growth and bidirectional coverage-\defect relationship highlight the need for diff-aware fuzzing techniques that effectively target modified code regions.

Future research should investigate \defect types under different conditions and develop more sophisticated fuzzing strategies for continuous integration environments.







\section*{Acknowledgment}
We gratefully acknowledge the financial support of JSPS for the KAKENHI grants (JP24K02921, JP24K02923, JP25K21359), as well as JST for the PRESTO grant (JPMJPR22P3), the ASPIRE grant (JPMJAP2415), and the AIP Accelerated Program (JPMJCR25U7).

\ifCLASSOPTIONcaptionsoff
  \newpage
\fi

\bibliographystyle{IEEEtran}

\bibliography{references}

\begin{IEEEbiography}[{\includegraphics[width=1in,height=1.25in,clip,keepaspectratio]{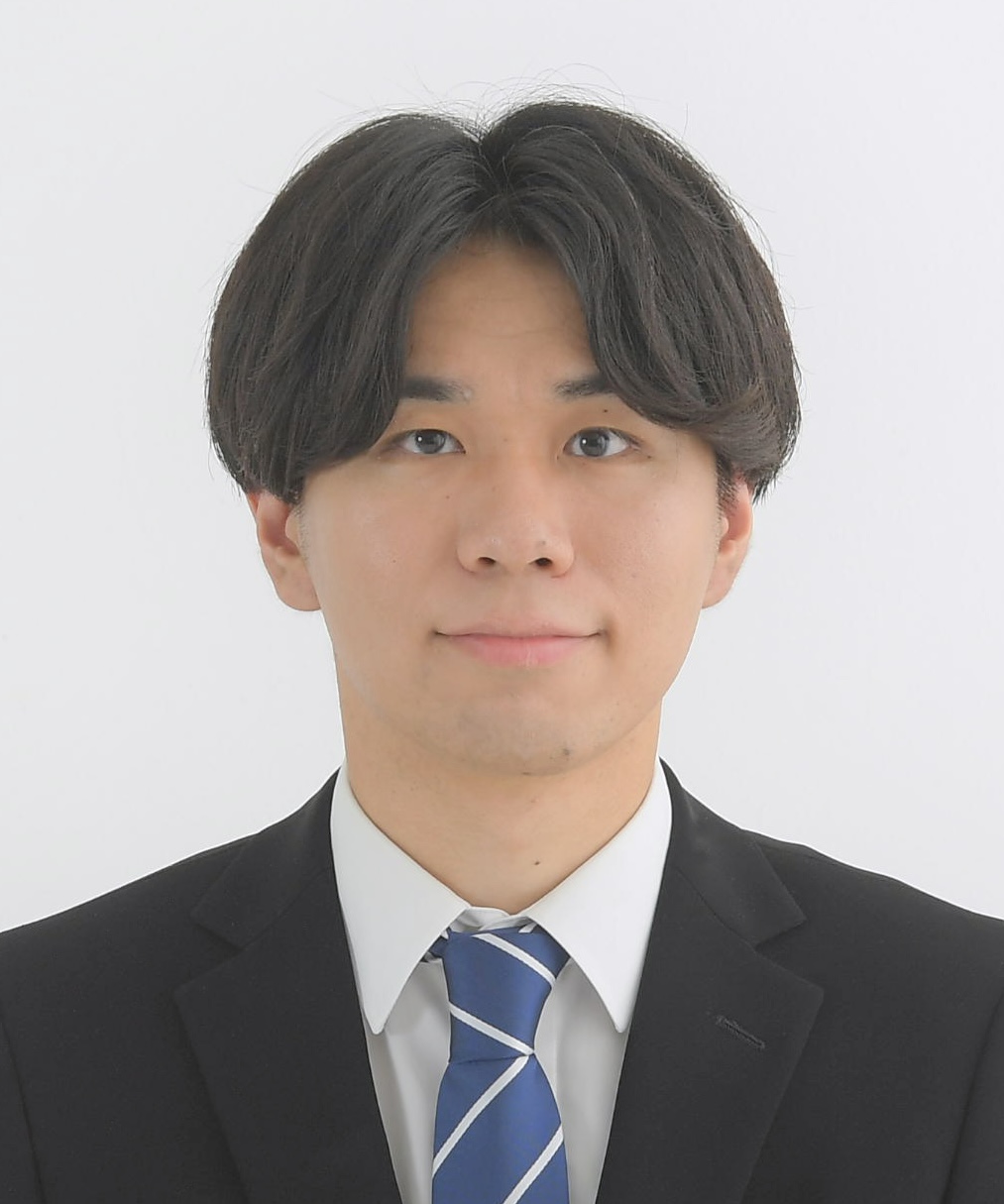}}]{Tasuya Shirai} is currently a Master's student at the Graduate School of Science and Technology, Nara Institute of Science and Technology (NAIST). He received the Bachelor of Engineering degree from Kumamoto National College of Technology in 2024. He is currently conducting research toward a doctoral degree under the supervision of Dr. Yutaro Kashiwa and Dr. Hajimu Iida. His research interests include fuzzing, especially continuous fuzzing.
\end{IEEEbiography}

\begin{IEEEbiography}[{\includegraphics[width=1in,height=1.25in,clip,keepaspectratio]{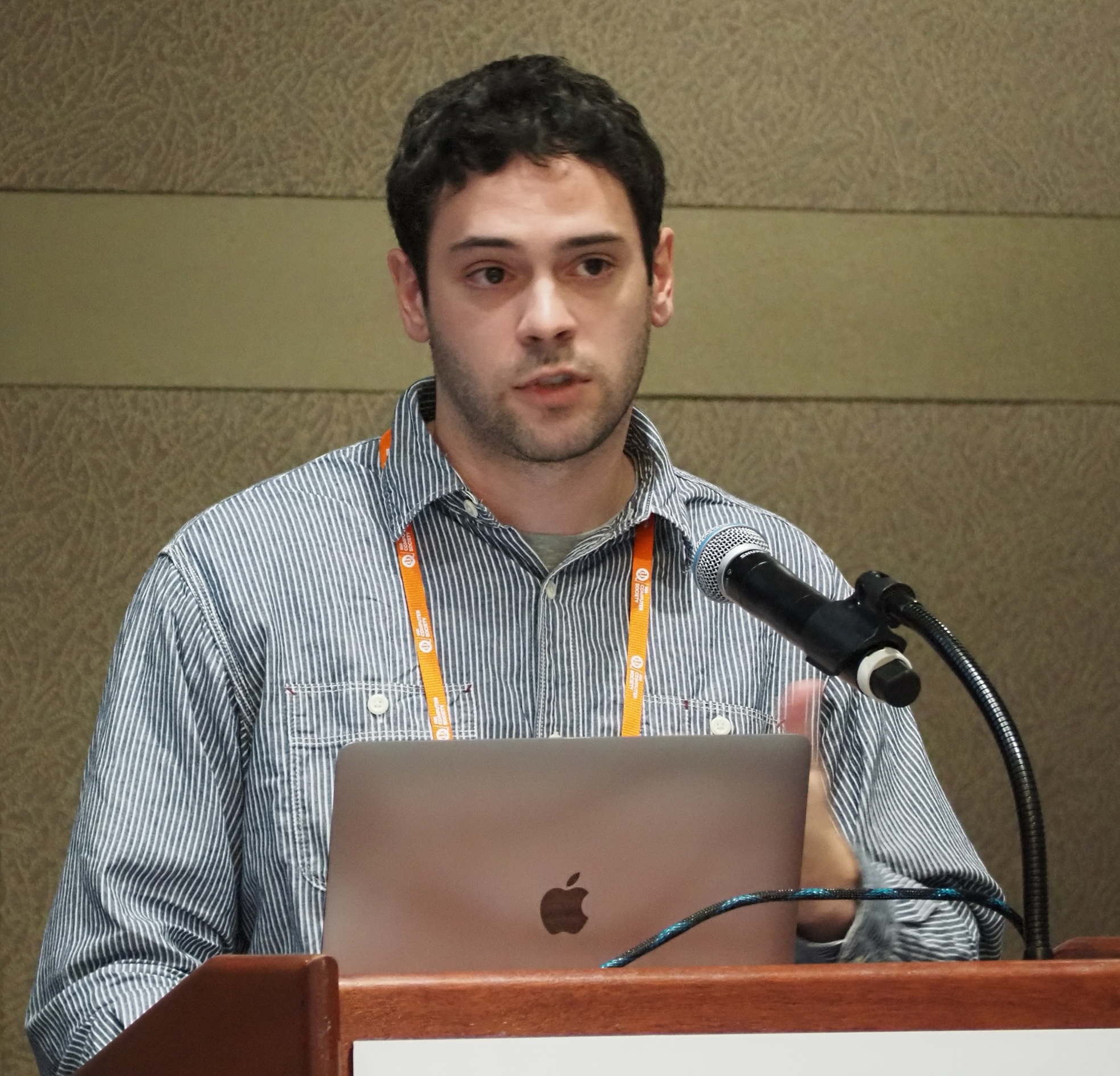}}]{Olivier Nourry}
is an assistant professor at Osaka University, Japan. He received his Ph.D. degree in Information Science and Technology from Kyushu University in 2024. His research interests include fuzzing, automated vulnerability repair, and empirical software engineering. He has many publications regarding continuous fuzzing in major journals and conferences such as TOSEM and ICSME.
\end{IEEEbiography}


\begin{IEEEbiography}[{\includegraphics[width=1in,height=1.25in,clip,keepaspectratio]{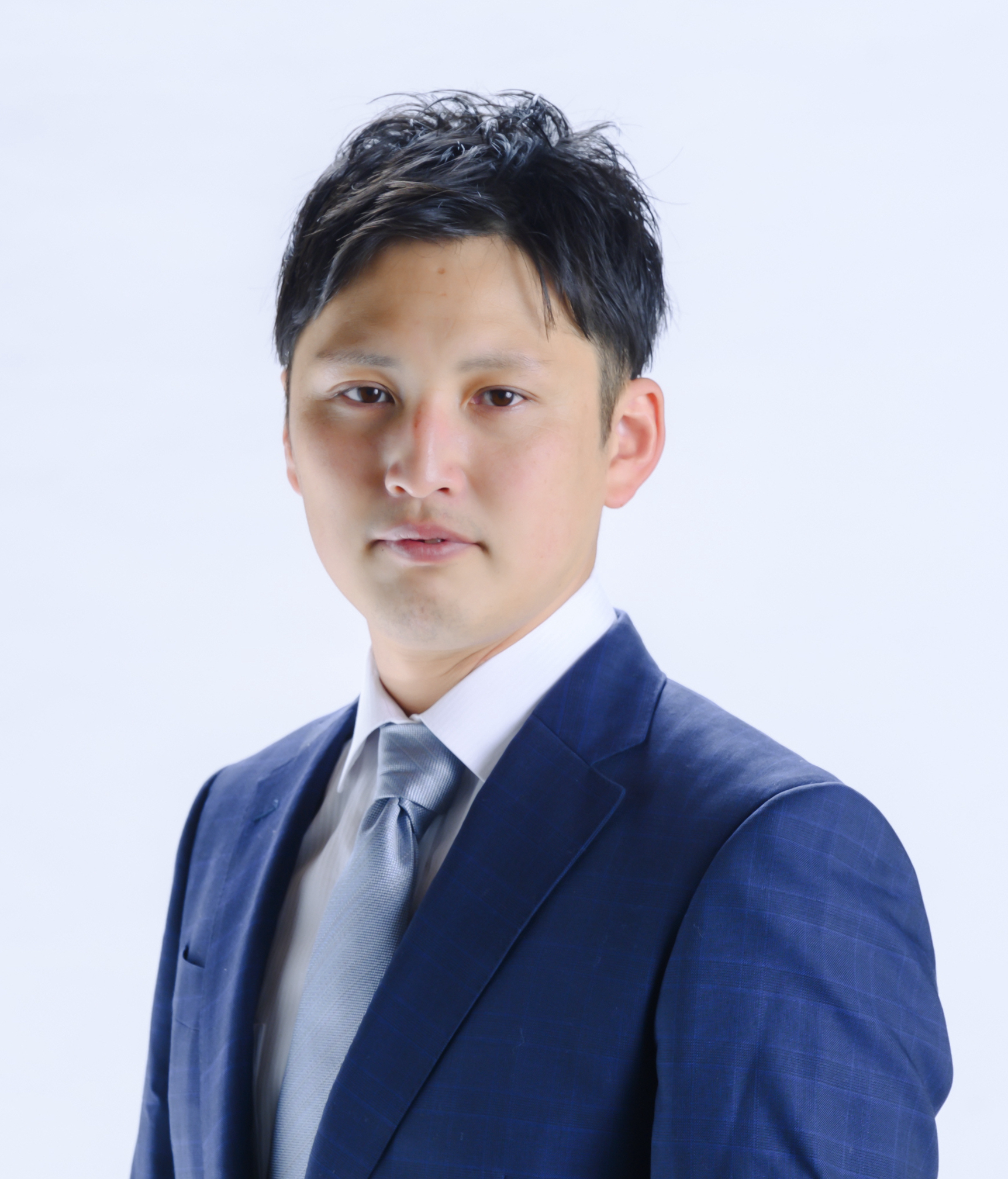}}]{Yutaro Kashiwa}
is an associate professor at Nara Institute of Science and Technology (NAIST), Japan. He worked for Hitachi Ltd. as a full-time software engineer for two years before spending three years as a research fellow of the Japan Society for the Promotion of Science. He received his Ph.D. degree in engineering from Wakayama University in 2020. His research interests include empirical software engineering, specifically the analysis of testing, refactoring, and releasing.
\end{IEEEbiography}

\begin{IEEEbiography}
[{\includegraphics[width=1in,height=1.25in,clip,keepaspectratio]{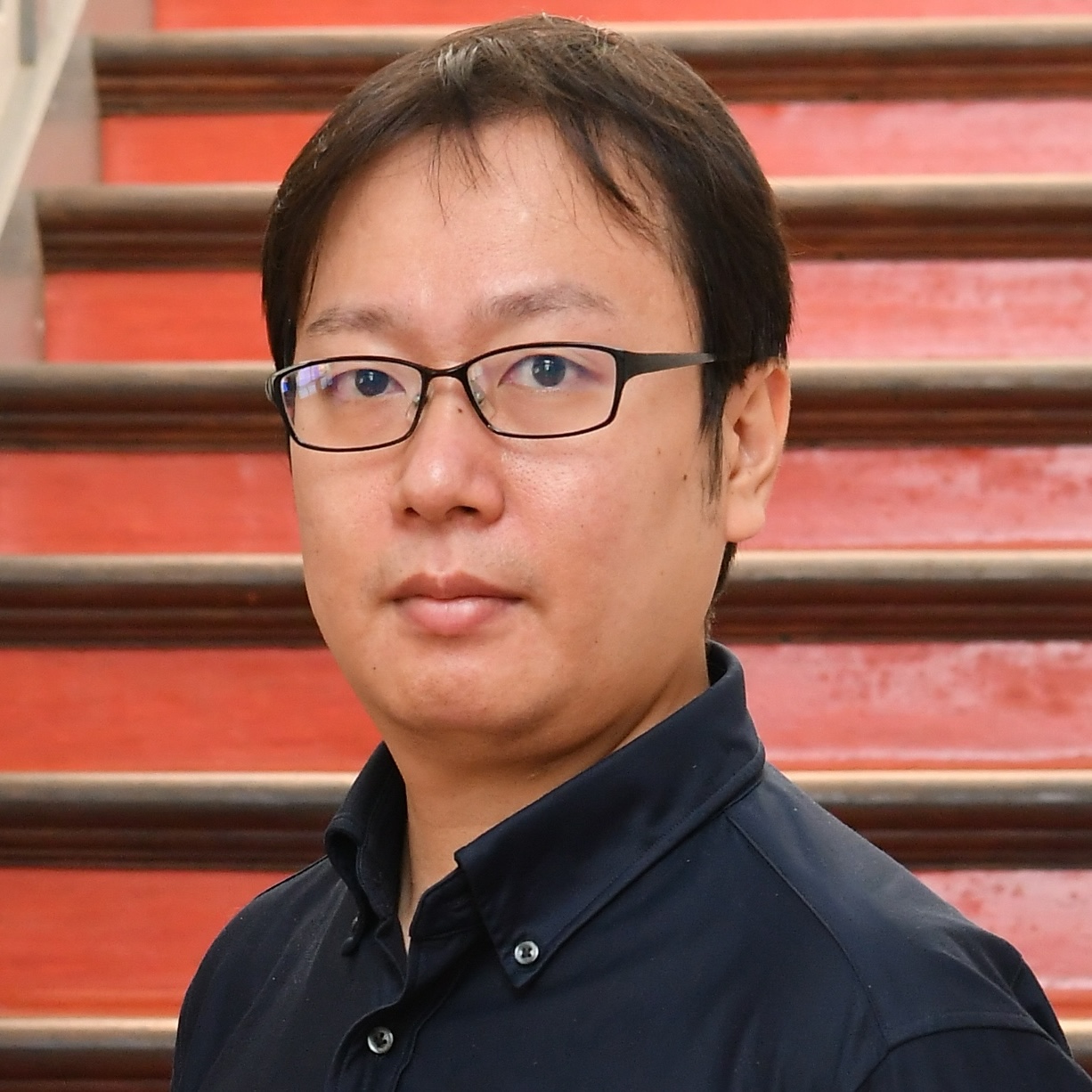}}]
{Kenji Fujiwara}
received the Ph.D. degree from the Nara Institute of Science and Technology in 2015. He is currently a Lecturer at Nara Women’s University, Japan. 
Before joining Nara Women’s University, he was a Lecturer at Tokyo City University, from 2021 to 2023, and an Assistant Professor at National Institute of Technology, Toyota College, from 2016 to 2021. 
His research interests include software maintenance and testing, and support for programming education.
\end{IEEEbiography}

\begin{IEEEbiography}[{\includegraphics[width=1in,height=1.25in,clip,keepaspectratio]{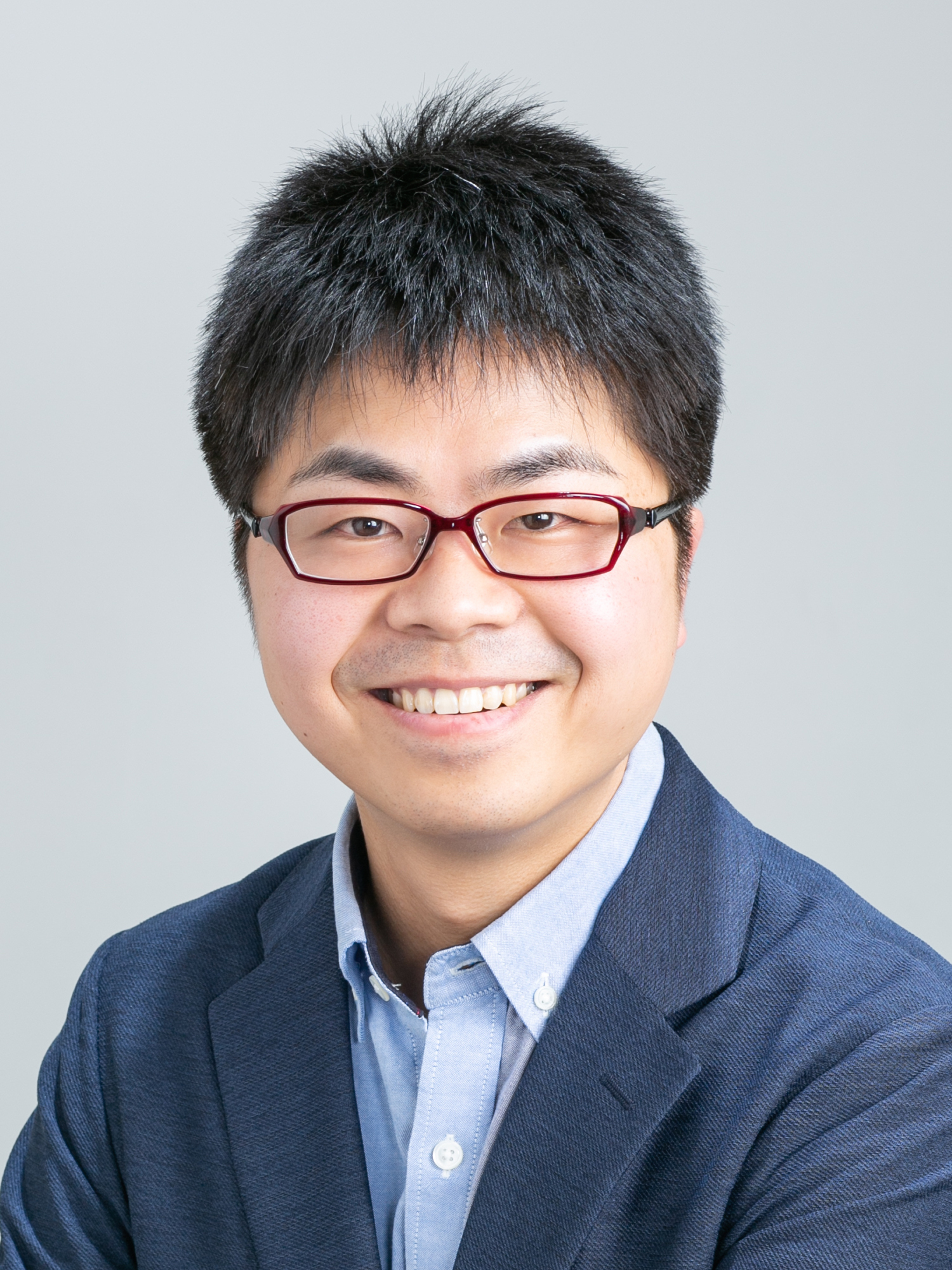}}]{Yasutaka Kamei} is a full professor at Kyushu University in Japan. He was a research fellow of the JSPS (PD) and was a postdoctoral fellow at Queen’s University in Canada. He received Ph.D. degree in Information Science from Nara Institute of Science and Technology. His research interests include empirical software engineering, open source software engineering and Mining Software Repositories (MSR). He served as a program-committee co-chair of SANER 2016 and MSR 2018 conferences. 
\end{IEEEbiography}

\begin{IEEEbiography}[{\includegraphics[width=1in,height=1.25in,clip,keepaspectratio]{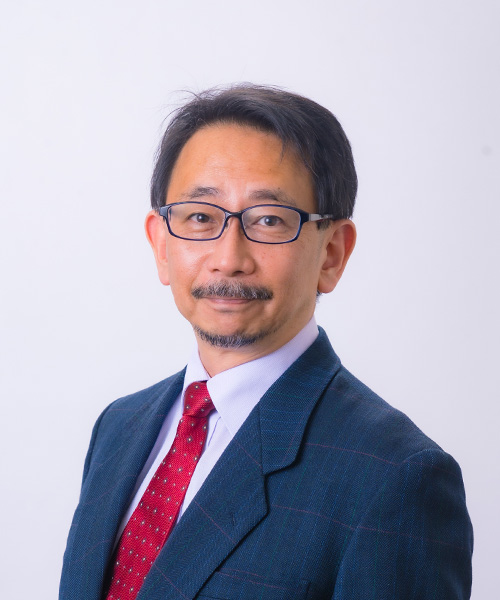}}]{Hajimu Iida}
 is a full professor at the Graduate School of Science and Technology, Nara Institute of Science and Technology (NAIST), Japan, where he leads the Laboratory for Software Design and Analysis. He received his Ph.D. in Engineering from Osaka University in 1994. His research interests span a wide range of topics in software engineering, including software process modeling and analysis, quantitative project management, repository mining, and software analytics. 
\end{IEEEbiography}
~\\
~\\
~\\
~\\
~\\
~\\
~\\
~\\
~\\
~\\
~\\
~\\
\end{document}